\documentclass[aps,prd,showpacs,article]{revtex4}
\usepackage{amsmath}
\usepackage{graphicx}
\usepackage{epsfig}

\begin{document}

\title{Renormalized Wick expansion for a modified PQCD}
\author{Alejandro Cabo Montes de Oca}
\affiliation{\ International Center for Theoretical Physics,
Strada Costiera 111, Trieste, Italy }

\affiliation{\ Group of Theoretical Physics, Instituto de
Cibern\'etica, Matem\'atica y F\'{\i}sica, Calle E, No. 309,
Vedado, La Habana, Cuba}
\begin{abstract}
\noindent The renormalization scheme for the Wick expansion of a modified
version of the perturbative QCD introduced in previous works is discussed.
Massless QCD is considered, by implementing the usual multiplicative scaling
of the gluon and quark wave functions and vertices. However, also massive
quark and gluon counter-terms are allowed in this mass less theory since the
condensates are expected to generate masses. A natural set of expansion
parameters of the physical quantities is introduced: the coupling itself and
to masses $m_q$ and $m_g$ associated to quarks and gluons respectively. This
procedure allows to implement a dimensional transmutation effect through
these new mass scales. A general expression for the new generating
functional in terms of the mass parameters $m_q$ and $m_g$ is obtained in
terms of integrals over arbitrary but constant gluon or quark fields in each
case. Further, the one loop potential, is evaluated in more detail in the
case when only the quark condensate is retained. This lowest order result
again indicates the dynamical generation of quark condensates in the vacuum.
\end{abstract}
 \maketitle

\section{Introduction}

The relevance of properly understanding QCD is difficult to overestimate.
However, due to the known difficulties associated to its strong interaction
properties, the predictions of QCD are also extremely far from a
satisfactory knowledge. Thus, the investigation of the properties of the
theory should be attacked from all possible angles, as it has been
undertaken along many years \cite
{hadron1,hadron2,fritzsch,savvidi,cjt,mirans,cornwall1,bardeen,shabad}. The
motivations for considering this work (and a few of previous ones done in
the theme in conjunction with other colleagues \cite{mpla,
prd,epjc,hoyer,epjc1,epjc2,hoyer1,hoyer2,jhep}) can be resumed as follows:
Firstly, to consider that within modern views in high energy physics, the
masses are normally searched to appear as generated by a spontaneous or
dynamical symmetry breaking in starting mass less theories. Then, the
circumstance that the QCD conveys the strongest forces of Nature, in
combination with the fact that the mass less version of the theory has not a
definite mass parameter, directly leads to the physical relevance of
examining the scales of dynamical mass allowed by symmetry breaking
processes in mass less QCD \cite{coleman,mirans}. In former works we have
got indications about the possibility of generating masses in QCD \cite
{epjc,jhep}. In \cite{epjc} modified Feynman rules were employed for
evaluating the quark masses from the Dyson equation in which the simplest
corrections to the self-energy determined by the condensates were retained.
For this purpose the gluon condensate parameter $C_g$ was evaluated by
fixing the mean value of the gluon Lagrangian (a quantity which in the
proposed picture is non-vanishing in the lowest approximation) to its
estimated value in the literature. The result for initially massless quarks,
surprisingly gave a value of one third of the proton mass \cite{epjc}. That
is, a prediction of the constituent quark masses followed. \newline

Motivated by this result, in Ref. \cite{jhep} we considered similar
evaluations assuming also the presence of a quark condensates for any
flavour $C_f$, $f=1,2...6$ and for the gluons $C_g$. In this case, by
properly selecting the coefficients $C_f$ and $C_g$, it was possible to
obtain the quark masses as singularities of the propagator for the six
quarks, also in the simplest approximation. Thus, the question emerged about
the possibility for those condensate values to be generated as the result of
a dynamical symmetry breaking in mass less QCD. In Ref. \cite{epjc2} this
issue was started to be considered by evaluating particular summations of
one loop diagrams. The results were positive in the direction of supporting
the generation of the gluon as well as the quark condensates. However, the
instability of the potential at zero value signaling the production of the
quark condensate, had not a bounded from below form. This fact did not
allowed to predict a concrete mean value of the condensate to be approached
by the system after stabilization. However, this result could be a
consequence of $\ $first evaluations in a scheme in which the
renormalization was not yet implemented or of the low order corrections
which were calculated. \newline

Therefore, in this work we start considering the renormalization of the
proposed expansion and its application to evaluate the first order
contributions to the Effective Action in terms of the quark and gluon
condensates. For this purpose, we are already supported by the discussion in
Refs. \cite{epjc1} in which the gauge invariance of the proposed Feynman
expansion was argued. Also and importantly, a procedure was devised for
eliminating the apparent singularities appearing in the graph expansion due
to the presence of Dirac Delta functions of the momenta \cite{epjc1,capper}.

Here we start by introducing the renormalization prescription in the
Euclidean version of the generating function of the Green functions $Z=\exp
(W)$ depending of the external sources, which also will be a function of the
condensate parameters for quarks and gluons. Multiplicative renormalization
is implemented for the fields and coupling constants in the Wick expansion.
However, it should be noticed that mass counter-terms will be also added,
although the bare theory before the adiabatic connection of the interaction
is being assumed mass less QCD. This assumption is essential: we are
considering that the mass counter-terms will be automatically generated by
the interactions, although they can not be implemented by the multiplicative
scaling of the fields and parameters. The Effective Action is then
introduced as usual, as the Legendre transform over the external sources in
favour of the mean values of the quantum fields. The effective potential
determined by it, is also a function of the gluon and quark condensates and
naturally it should show a minimum with respect to these quantities at the
ground state. Here, the fields values are fixed to vanish in the ground
state assuming the Lorentz invariance of the vacuum from the start. All the
generating functionals defined are considered as expanded in power series of
the coupling constant $g$, and the two defined mass parameters $%
m_q=(g^2C_q)^{\frac 13}$ and $m_g=(g^2C_g)^{\frac 12}$. This reordering of
the expansion of the physical quantities allows to implement the dimensional
transmutation effect in the considered mass less QCD. The generating
functional $Z$ is obtained in a form that resumes the effect of the
condensates as generated by Gaussian weighted averages over constant
homogeneous background gauge fields for gluons and quarks, respectively.
These formulae are expected to be considered for detailed calculations
elsewhere. Finally, the lowest order correction to the potential is
evaluated in more detail for the case of the single presence of the quark
condensate. The dependence of the potential indicates a tendency to the
dynamical generation of the condensate in the lowest order.

The work will proceed as follows. In Section 2, the Feynman expansion for
QCD in Euclidean variables and the conventions to be used, will be
described. Next, in the Section the renormalized generating functional
expressing the Feynman expansion in momentum space is written. Section 3
considers the calculation of the Effective Potential in zero order in $g$
for the case of the single presence of the quark condensate. Finally, the
results are reviewed and commented in the Summary.

\section{ Generating functional $Z[j,\eta ,\overline{\eta },\xi \overline{%
,\xi }]$}

As remarked in the Introduction the purpose of the this Section is to
present a renormalized version of the proposed perturbative expansion. For
this aim the starting action for mass less QCD in Euclidean variables will
be taken in the form
\begin{equation}
S=\int dx(-\frac 14F_{\mu \nu }^aF_{\mu \nu }^a-\frac 1{2\alpha }\partial
_\mu A_\mu ^a\partial _\nu A_\nu ^a-\overline{\Psi ^i}_q\text{ }i\gamma _\mu
D_\mu ^{ij}\Psi _q^j-\overline{c}^a\partial _\mu D_\mu ^{ab}c^b),\text{ \ }
\end{equation}
where the field intensity, and covariant derivatives follow the conventions
\begin{eqnarray}
F_{\mu \nu }^a &=&\partial _\mu A_\nu ^a-\partial _\nu A_\mu
^a-gf^{abc}A_\mu ^bA_\nu ^c,  \nonumber \\
D_\mu ^{ij} &=&\partial _\mu \delta ^{ij}+ig\text{ }A_\mu ^aT_a^{ij},\ \ \ \
D_\mu ^{ab}=\partial _\mu \delta ^{ab}+gf^{abc}\text{ }A_\mu ^c, \\
\{\gamma _\mu ,\gamma _\nu \} &=&-2\delta _{\mu \nu },\ \ \ \
[T_aT_b]=if^{abc}T_c.  \nonumber
\end{eqnarray}

The Fourier decomposition for any field, i.e. the gauge one, will assumed in
the form
\begin{eqnarray*}
A_\mu ^b(x) &=&\int \frac{dk}{(2\pi )^D}\exp (ik_\mu \text{\thinspace }x_\mu
)\text{ }A_\mu ^b(k), \\
A_\mu ^b(k) &=&\int dx\,\exp (-ik_\mu \text{\thinspace }x_\mu )\text{ }A_\mu
^b(x).
\end{eqnarray*}

Then, the renormalized Green's functions generating functional including the
gluon and quark condensate parameters $C_q$ and $C_g$ as discussed in Ref.
\cite{epjc1}, is expressed as follows
\begin{eqnarray}
Z[j,\eta ,\overline{\eta },\xi ,\overline{\xi }] &=&\frac{I[j,\eta ,%
\overline{\eta },\xi \overline{,\xi }]}{I[0,0,0,0]},  \nonumber \\
I[j,\eta ,\overline{\eta },\xi ,\overline{\xi }] &=&\exp (V^{int}[\frac
\delta {\delta j},\frac \delta {\delta \overline{\eta }},\frac \delta
{-\delta \eta },\frac \delta {\delta \overline{\xi }}\frac \delta {-\delta
\xi }]){\small \times }  \label{Wick} \\
&&\exp (\int \frac{dk}{(2\pi )^D}j(-k)\frac 12D(k)j(k)){\small \times }
\nonumber \\
&&\exp (\int \frac{dk}{(2\pi )^D}\overline{\eta }(-k)G_q(k)\eta (k)){\small %
\times }  \nonumber \\
&&\exp (\int \frac{dk}{(2\pi )^D}\overline{\xi }(-k)G_{gh}(k)\xi (k)){\small %
,}  \nonumber
\end{eqnarray}
in which the only changes with respect to the functional associated to the
usual perturbative QCD\ appear in only two of the three free propagators of
the expansion, the quark and the gluon ones\cite{mpla,
prd,epjc,hoyer,epjc1,epjc2,hoyer1,hoyer2,jhep}:
\begin{eqnarray}
D_{\mu \nu }^{ab}(k) &=&\delta ^{ab}(\frac 1{k^2}(\delta _{\mu \nu }-\frac{%
k_\mu k_\nu }{k^2})\theta _N(k)+C_g^b\delta ^D(k)\delta _{\mu \nu }),
\nonumber \\
G_f^{ij}(k) &=&\delta ^{ij}(\frac{\theta _N(k)}{m_f+\gamma _\mu k_\mu }%
+C_q^b\delta ^D(k)I),  \nonumber \\
G_{gh}^{ab}(k) &=&\delta ^{ab}\frac{\theta _N(k)}{k^2}.  \label{propagators}
\end{eqnarray}

That is, the quark and gluon propagators now include the condensation
effects through the $C_q$ and $C_g$ parameters respectively. However, an
important additional change is also present in (\ref{propagators}). The
usual Feynman propagators are regularized in the neighborhood of zero
momentum by multiplying them by
\begin{equation}
\theta _N(k)=\theta (\sigma -|k|),\;|k|=(k_\mu k_\mu )^{\frac 12},
\end{equation}
where $\theta $ is the Heaviside function and $\sigma $ is an infinitesimal
momenta cutoff, that we will call as the Nakanishi parameter. As discussed
in \cite{epjc1}, this regularization naturally arise when gauge theory
propagators are properly defined. Here we assumed that upon passing to the
Euclidean version of the theory, the prescription can be translated to
eliminate a neighborhood of the momentum space near the zero momentum point.
This infrared regularization in combination with a dimensional
regularization rule for products of Dirac Delta functions evaluated at zero
momenta, allowed in Ref. \cite{epjc1} to eliminate the singularities
appearing in the perturbative series by the presence of products of Delta
functions evaluated at zero momenta and of them with propagators evaluated
also at vanishing momenta.

The vertices defining the interaction in the Wick expansion formula (\ref
{Wick}) have the decomposition
\begin{eqnarray}
V^{int} &=&V_g^{(1)}+V_g^{(2)}+V_q^{(1)}+V_{gh}^{(1)}+  \nonumber \\
&&(Z_1-1)V_g^{(1)}+(Z_4-1)V_g^{(2)}+(Z_{1F}-1)V_q^{(1)}+  \nonumber \\
&&(\widetilde{Z}_1-1)V_{gh}^{(1)}+  \label{vint} \\
&&V_g^{(0)}+V_q^{(0)}+V_{gh}^{(0)},  \nonumber
\end{eqnarray}
where the superindices indicate the order the coupling constant associated
to the original vertices in the bare action. As usual the unrenormalized
bare fields (signaled with a superindex $b)$ will be related to their
renormalized counterparts through the factors $"Z\,"$ as
\begin{eqnarray}
A_\mu ^b &=&Z_3^{\frac 12}A_\mu ,  \nonumber \\
\Psi _q^b &=&Z_2^{\frac 12}\Psi _q,\;\overline{\Psi }_q^b=Z_2^{\frac 12}%
\overline{\Psi }_q, \\
\chi ^b &=&Z_3^{\frac 12}\chi ,\;\;\;\overline{\chi }^b=Z_3^{\frac 12}%
\overline{\chi },  \nonumber
\end{eqnarray}
with the also usual correspondence between the sources
\begin{eqnarray}
j_\mu ^b &=&Z_3^{-\frac 12}j_\mu ,  \nonumber \\
\eta _q^b &=&Z_2^{-\frac 12}\eta _q,\text{ \ }\overline{\eta }%
_q^b=Z_2^{-\frac 12}\overline{\eta }_q, \\
\xi ^b &=&Z_3^{-\frac 12}\xi ,\text{ \ \ }\overline{\xi }^b=Z_3^{-\frac 12}%
\overline{\xi }.  \nonumber
\end{eqnarray}
It should be underlined that the bare condensate parameters $C_q^b$ and $%
C_g^b$ appear in the free propagators because, these constants had not been
expanded yet in their renormalized and counterterm contributions. This will
be performed later within what we think is a more convenient representation
for this purpose. It is also assumed that the scale parameter $\mu $ of
dimensional regularization links the dimensional coupling $g$ with its
dimensionless value $g_o$ as
\begin{equation}
g=g_0\mu ^{2-\frac D2}=g_0\mu ^\epsilon .
\end{equation}
The expressions for each of the vertices entering $V^{int}$ in (\ref{vint})
are given in the Appendix A. Let us examine more closely the vertex $%
V_q^{(0)}$ associated to the mass and wave function renormalization of the
bare theory. It should be first recalled that the modified expansion under
consideration was motivated by deriving the Wick expansion for a mass less
QCD in which the mass parameter is absent. Then, the renormalization
procedure being investigated is consequently assumed to represent the
physics of an adiabatic connection of the interaction from an originally
mass less theory. Therefore, we estimate as the most natural procedure to
fix the bare masses of gluons and quarks as vanishing. However, it is clear
that since the theory has been argued to generate mass \cite{epjc,jhep}, the
connection of the interaction should be expected to produce mass
counterterms in the renormalized action. These terms will also assumed to
appear among the quark counterterm vertices $V_q^{(0)}$ in Appendix A with
the form:
\begin{eqnarray*}
V_q^{(0)}[\frac \delta {\delta \overline{\eta }},\frac \delta {-\delta \eta
}] &=&\int \int dk_1dk_2(2\pi )^D\delta ^D(k_1+k_2)\times \\
&&\frac \delta {\delta \eta (k_2)}\left( (Z_2-1)k_{1\mu }\gamma _\mu +\delta
m(g_0,m_q,m_g)\right) \frac \delta {\delta \overline{\eta }(k_1)},
\end{eqnarray*}
although a non vanishing $\delta m$ mean a break of pure multiplicative
renormalization. This seems to be not a complication since multiplicative;
renormalization is known to be broken when the theory has no symmetries that
enforces the vanishing of allowed counterterms having no counterpart in the
original bare action.

\strut Now, the source terms in the gluon condensate propagator can be
represented as a Gaussian integral over constant and homogeneous gauge boson
fields as follows
\begin{eqnarray}
\exp [\left( \int \frac{dk}{(2\pi )^D}j_\mu ^a(-k)C_g^b\delta (k)j_\mu
^a(k)\right) ] &=&\exp \left( \frac{C_g^b}{(2\pi )^D}j_\mu ^a(0)j_\mu
^a(0)\right) ,  \nonumber \\
&=&\frac 1{(2\pi )^{(N^2-1)D}\mathcal{N}_g\mathcal{(}Z_3\mathcal{)}}\int
d\alpha _\mu ^a\exp [(\frac{Z_3-1}{Z_3})\frac{\alpha _\mu ^a\alpha _\mu ^a}%
2]\times   \nonumber \\
&&\exp [-\frac{\alpha _\mu ^a\alpha _\mu ^a}2+(\frac{2C_g}{(2\pi )^D}%
)^{\frac 12}j_\mu ^a(0)\alpha _\mu ^a],  \label{form1} \\
&=&\frac 1{(2\pi )^{(N^2-1)D}\mathcal{N}_g\mathcal{(}Z_3\mathcal{)}}\exp [%
\frac{(Z_3-1)}2(\frac{2C_g}{(2\pi )^D})^{-1}\frac{\partial ^2}{2\partial
j_\mu ^a\partial j_\mu ^a}]\times   \nonumber \\
&&\int d\alpha _\mu ^a\exp \left( -\frac{\alpha _\mu ^a\alpha _\mu ^a}2+(%
\frac{2C_g}{(2\pi )^D})^{\frac 12}j_\mu ^a(0)\alpha _\mu ^a\right) ,
\nonumber
\end{eqnarray}
where $\mathcal{N}_g$ is a normalization cons$\tan $t which cancels with a
similar one appearing in the normalizing factor $I[0,0,0,0]$ in (\ref{Wick}%
). Here,the process of, let say, ''translating'' the renormalization part $%
(Z_3-1)C_g$ of the gluon condensate to the counterterms is started by
expressing the terms containing those parts as an exponential of a quadratic
form in the derivatives over the sources.

For quarks, an analog formula in terms of interaction over anti-commuting
fermion fields will be employed. It has the form
\begin{eqnarray}
\exp \left( \int \frac{dk}{(2\pi )^D}\overline{\eta }_u^i(-k)C_q^b\delta
(k)\eta _u^i(k)\right)  &=&\exp \left( \frac{C_q^b}{(2\pi )^D}\overline{\eta
}_u^i(0)\eta _u^i(0)\right)   \nonumber \\
&=&\frac 1{\mathcal{N}_q\mathcal{(}Z_2\mathcal{)}}\int d\overline{\chi }%
_u^id\chi _u^i\exp [\frac{(Z_2-1)}{Z_2}\overline{\chi }_u^i\chi _u^i]
\nonumber \\
&&\exp [-\overline{\chi }_u^i\chi _u^i+(\frac{C_q}{(2\pi )^D})^{\frac 12}(%
\overline{\eta }_u^i(0)\chi _u^i+\overline{\chi }_u^i\eta _u^i(0))],
\label{form2} \\
&=&\frac 1{\mathcal{N}_q\mathcal{(}Z_2\mathcal{)}}\exp [\frac{(Z_2-1)}{Z_2}(%
\frac{C_q}{(2\pi )^D})^{-1}\frac{\partial ^2}{-\partial \eta _u^i(0)\partial
\overline{\eta }_u^i(0)}]\times   \nonumber \\
&&\int d\overline{\chi }_u^id\chi _u^i\exp \left( -\frac{\overline{\chi }%
_u^i\chi _u^i}2+(\frac{C_q}{(2\pi )^D})^{\frac 12}(\overline{\eta }%
_u^i(0)\chi _u^i+\overline{\chi }_u^i\eta _u^i(0))\right) ,  \nonumber
\end{eqnarray}
in which again the appearing factor $\mathcal{N}_q$ will be cancelled by a
similar one appearing in $I[0,0,0,0]$. In the above expressions the
condensate parameters have been naturally decomposed in the way
\begin{eqnarray*}
C_g^b &=&Z_3C_g=(Z_3-1)C_g+C_g, \\
Cq &=&Z_2C_q=(Z_2-1)C_g+C_g.
\end{eqnarray*}
It can be recalled that the condensates were created by the action over the
vacuum of exponential of quadratic forms for the bare gluon and quark
creation operators \cite{prd,epjc1}. Therefore, the renormalization of these
operators implies that the bare parameters should be related to the
renormalized ones through the same $Z_3$ or $Z_2$ constants for gluons and
quarks respectively. It can be noted that within the quadratic form defining
the gluon condensate quadratic terms in the ghost fields also appeared.
However, since the ghosts have the same renormalization constant that the
gluons, the $Z_3$ proportionality between the bare and renormalized gluon
condensates should remain valid.

The following relationships help to transform the functional derivatives
over the sources, when integrated around zero momentum within the Nakanishi
neighborhood, as usual derivatives over the zero momentum components of
these sources.
\begin{eqnarray*}
\int dk\frac \delta {\delta j(k)}F[j,\eta ,\overline{\eta },\xi ,\overline{%
\xi }]\,\theta _N(k) &=&\frac \partial {\partial j(0)}F[j,\eta ,\overline{%
\eta },\xi ,\overline{\xi }], \\
\int dk\frac \delta {\delta \eta (k)}F[j,\eta ,\overline{\eta },\xi ,%
\overline{\xi }]\,\theta _N(k) &=&\frac \partial {\partial \eta (0)}F[j,\eta
,\overline{\eta },\xi ,\overline{\xi }], \\
\int dk\frac \delta {\delta \overline{\eta }(k)}F[j,\eta ,\overline{\eta }%
,\xi ,\overline{\xi }]\,\theta _N(k) &=&\frac \partial {\partial \overline{%
\eta }(0)}F[j,\eta ,\overline{\eta },\xi ,\overline{\xi }], \\
\int dk\frac \delta {\delta \xi (k)}F[j,\eta ,\overline{\eta },\xi ,%
\overline{\xi }]\,\theta _N(k) &=&0, \\
\int dk\frac \delta {\delta \overline{\xi }(k)}F[j,\eta ,\overline{\eta }%
,\xi ,\overline{\xi }]\,\theta _N(k) &=&0.
\end{eqnarray*}

After employing the above relations, the $Z$ functional can be represented
in a form where the effects of the condensates are conveyed by the newly
incorporated gluon and quark constant and homogeneous fields $\alpha ,%
\overline{\chi }$ and $\chi $ (below, they will be named as the $auxiliary$ $%
fields).$ The expression for $Z$ is
\begin{eqnarray}
Z[j,\eta ,\overline{\eta },\xi \overline{,\xi }] &=&\frac{I[j,\eta ,%
\overline{\eta },\xi \overline{,\xi }]}{I[0,0,0,0]},  \nonumber \\
I[j,\eta ,\overline{\eta },\xi ,\overline{\xi }] &=&\frac 1{\mathcal{N}}\int
\int d\alpha d\overline{\chi }d\chi \exp [-\overline{\chi }_u^i\chi _u^i-%
\frac{\alpha _\mu ^a\alpha _\mu ^a}2]  \nonumber \\
&&\exp {\small [}\widehat{V}^{int}{\small [}\frac \delta {\delta j}{\small +(%
}\frac{2C_g}{(2\pi )^D}{\small )}^{\frac 12}{\small \alpha ,}\frac \delta
{\delta \overline{\eta }}{\small +(}\frac{C_q}{(2\pi )^D}{\small )}^{\frac
12}{\small \chi ,}\frac \delta {-\delta \eta }{\small +(}\frac{C_q}{(2\pi )^D%
}{\small )}^{\frac 12}\overline{\chi }{\small ,}\frac \delta {\delta
\overline{\xi }}\frac \delta {-\delta \xi }{\small ,\alpha ,}\overline{\chi }%
{\small ,\chi ]]\times }  \label{shifted} \\
&&\exp {\small [}\int \frac{dk}{(2\pi )^D}{\small j(-k)}\frac 12{\small D}^F%
{\small (k)j(k)]\times }  \nonumber \\
&&\exp {\small [}\int \frac{dk}{(2\pi )^D}\overline{\eta }{\small (-k)G}_q^F%
{\small (k)\eta (k)]\times }  \nonumber \\
&&\exp {\small [}\int \frac{dk}{(2\pi )^D}\overline{\xi }{\small (-k)G}%
_{gh}^F{\small (k)\xi (k)].}  \nonumber
\end{eqnarray}
in which $\mathcal{N}$ is a normalization constant that again cancels in the
cocient of $I$ functions in (\ref{Wick}). Note that the propagators now are
the usual mass less Feynman ones, but the vertex terms $\widehat{V}^{int}$
include additional contributions associated to the renormalization of the
condensate parameters. The vertices expressed in terms of the derivatives
over the sources have the form
\begin{eqnarray*}
&&\widehat{V}^{int}[\frac \delta {\delta j},\frac \delta {\delta \overline{%
\eta }},\frac \delta {-\delta \eta },\frac \delta {\delta \overline{\xi }%
},\frac \delta {-\delta \xi },\frac \partial {\partial j(0)},\frac \partial
{-\partial \eta (0)},\frac \partial {\partial \overline{\eta }(0)}] \\
&=&V^{int}[\frac \delta {\delta j},\frac \delta {\delta \overline{\eta }%
},\frac \delta {-\delta \eta },\frac \delta {\delta \overline{\xi }}\frac
\delta {-\delta \xi }]+ \\
&&+\frac{(Z_3-1)}{Z_3}(\frac{2C_g}{(2\pi )^D})^{-1}\frac{\partial ^2}{%
2\partial j_\mu ^a(0)\partial j_\mu ^a(0)}] \\
&&+\frac{(Z_2-1)}{Z_2}(\frac{C_q}{(2\pi )^D})^{-1}\frac{\partial ^2}{%
-\partial \eta _u^i(0)\partial \overline{\eta }_u^i(0)}.
\end{eqnarray*}

Let us search now in what follows for a reordering of the perturbative
expansion seeking for explicitly introduce the dimensional transmutation
effect in the modified representation \cite{coleman}.

For this purpose a first idea comes from the fact that the auxiliary fields
enter as sorts of background constant fields. This fact directly leads to a
similar proposal to one made in Ref. \cite{mpla}. In that work, the modified
propagators considered here were first introduced for to be employed in the
modifying the perturbative expansion. However, there, it was also discussed
an alternative scheme in which the generating functional $Z$ for QCD was
considered as an average over constant gluon fields. This superposition
allowed to argue that the Fradkin's general functional differential
equations for $Z$ \cite{fradkin}, should be exactly obeyed by the mean value
over constant fields of auxiliary $Z$ functionals associated to arbitrary
constant mean fields. As it will be seen from the following discussion the
final form of the functional obtained here indicates the similarity between
the two proposals advanced in Ref. \cite{mpla}. However, in that work there
was not clarity about the possibility of \ introducing a weighted average,
and thus about how conveniently define it.

The auxiliary fields appear now in the vertices as kinds of constant gluon
or quark background fields. Therefore, it seems natural to express the
expansion (before the mean value over the auxiliary quantities is taken) in
terms of the gluon and quark propagators in the presence of such fields.

For this purpose, from $\widehat{V}^{int}$ in (\ref{shifted}) the
contribution to its expansion coming from the terms being second order in
the gluon and quark fields, but also including auxiliary backgrounds, will
be substracted. These terms, can be now acted on the exponential containing
the usual Feynman propagators contracted with the sources. Further, a
recourse can be employed of expressing back the exponential of the quadratic
forms of the sources in terms of the Feynman propagators in equation (\ref
{shifted}), as a continual integral over the gluon, quark and ghost fields.
Then, the previously mentioned exponential of the quadratic form in the
functional derivatives over the gluon and quark fields, simply will produce
an additional quadratic form within the exponential of the continual
integral. Collecting together the total quadratic form in the fields, leads
to a new modified free path integral over which the remaining exponential of
the vertices will act. Its expression is
\begin{eqnarray*}
Z^{(0)}[j,\eta ,\overline{\eta },\xi ,\overline{\xi }|C_q^b,C_g^b] &=&\frac
1{\mathcal{N}}\int \int d\alpha d\overline{\chi }d\chi \exp [-\overline{\chi
}_u^i\chi _u^i-\frac{\alpha _\mu ^a\alpha _\mu ^a}2]\int \mathcal{D}[A,%
\overline{\Psi },\Psi ,\overline{c},c]\times \\
&&\times \exp [-\int \frac{dk}{(2\pi )^D}[\frac 12A_\mu ^a(-k)(\mathbf{D}%
^{ac}\mathbf{D}^{cb}\delta _{\mu \nu }-\frac{\mathbf{D}_\mu ^{ac}\mathbf{D}%
_\nu ^{cb}\mathbf{+D}_\nu ^{ac}\mathbf{D}_\mu ^{cb}}2 \\
&&+\frac{\delta ^{ab}}\alpha k_\mu k_\nu )A_\nu ^b(k)+\int \frac{dk}{(2\pi
)^D}\overline{c}_\mu ^a(-k)k_\mu \mathbf{D}_\mu ^{ac}c(k) \\
&&+\int \frac{dk}{(2\pi )^D}\overline{\Psi }^i(-k)\text{ }\gamma _\mu
\mathbf{D}_\mu ^{ij}\,\Psi ^j(k)\text{ }+ \\
&&+\int \frac{dk}{(2\pi )^D}\overline{\Psi }^{i,u}(-k)\text{ }(-g)(\frac{%
C_q^b}{(2\pi )^D})^{\frac 12}\gamma _\mu ^{u\text{ }v}T_a^{ij}\,\chi ^{j,v}%
\text{ }A_\nu ^a(k)+ \\
&&+\int \frac{dk}{(2\pi )^D}A_\mu ^a(-k)\overline{\chi }^{i,u}\text{ }(-g)(%
\frac{C_q^b}{(2\pi )^D})^{\frac 12}\text{ }\gamma _\mu ^{u\text{ }%
v}T_a^{ij}\,\Psi ^{j,v}(k)+ \\
&&+\int \frac{dk}{(2\pi )^D}\overline{c}^ak^2c^b+\int \frac{dk}{(2\pi )^D}%
(j(-k)A(k)+\overline{\eta }(-k)\Psi (k)+ \\
&&\overline{\Psi }(k)\eta (-k)+\overline{\xi }(-k)c(k)+\overline{c}(-k)\xi
(k))],
\end{eqnarray*}
where $\mathbf{D}_\mu ^{ij}=k_\mu \delta ^{ij}+g(\frac{2Cg}{(2\pi )^D}%
)^{\frac 12}T_a^{ij}\alpha _\mu ^a$ and $\;\mathbf{D}_\mu ^{ab}=k_\mu \delta
^{ij}-ig(\frac{2Cg}{(2\pi )^D})^{\frac 12}f^{abc}\alpha _\mu ^c$ .

The above expression can be converted to a more compact form by defining a
composite field and its source and their conjugates, having boson and
fermion components, as follows
\begin{eqnarray}
\Phi &=&\left\{
\begin{array}{l}
A_\mu ^a \\
\Psi ^{r,u} \\
c^a
\end{array}
\right\} ,\Phi ^{*}=\overbrace{\underbrace{%
\begin{array}{lll}
A_\mu ^a & \overline{\Psi }^{r,u} & \overline{c}^a
\end{array}
}},  \label{sources} \\
J &=&\left\{
\begin{array}{l}
j_\mu ^a/2 \\
\eta ^{r,u} \\
\xi ^a
\end{array}
\right\} J^{*}=\overbrace{\underbrace{%
\begin{array}{lll}
j_\mu ^a/2 & \overline{\eta }^{r,u} & \overline{\xi }^a
\end{array}
}}.
\end{eqnarray}

Therefore, a new free generating functional $Z^{(0)}$ can then be expressed
in the way
\begin{eqnarray}
Z^{(0)}[j,\eta ,\overline{\eta },\xi \overline{,\xi }|C_q^b,C_g^b] &=&\frac
1{\mathcal{N}}\int d\alpha d\overline{\chi }d\chi \exp [-\overline{\chi }%
_u^i\chi _u^i-\frac{\alpha _\mu ^a\alpha _\mu ^a}2]\times \exp [-\frac
14V^DF_{\mu \nu }^a(\alpha )F_{\mu \nu }^a(\alpha )]  \nonumber \\
&&\int \mathcal{D}[\Phi ]\exp [\int \frac{dk}{(2\pi )^D}\Phi
^{*}(-k)S^{-1}(k)\Phi (k)+J^{*}(-k)\Phi (k)+\Phi ^{*}(-k)J(k)] \\
&=&\frac 1{\mathcal{N}}\int \int d\alpha d\overline{\chi }d\chi \exp [-%
\overline{\chi }_u^i\chi _u^i-\frac{\alpha _\mu ^a\alpha _\mu ^a}2]\times
\exp [-\frac 14V^DF_{\mu \nu }^a(\alpha )F_{\mu \nu }^a(\alpha )]  \label{Zo}
\\
&&\exp [\int \frac{dk}{(2\pi )^D}\Phi ^{*}(-k)S^{-1}(k)\Phi (k)]\exp
[J^{*}(-k)S(k)J(k)] \\
&=&\frac 1{\mathcal{N}}\int \int d\alpha d\overline{\chi }d\chi \exp [-%
\overline{\chi }_u^i\chi _u^i-\frac{\alpha _\mu ^a\alpha _\mu ^a}2]\mathcal{%
M(}\alpha ,\overline{\chi },\chi )\times   \nonumber \\
&&\exp [\int \frac{dk}{(2\pi )^D}J^{*}(-k)S(k)J(k)],
\end{eqnarray}
The quantity $F_{\mu \nu }^a(\alpha )=gf^{abc}\alpha _\mu ^b\alpha _\nu ^c$
is the field intensity of the gluon constant field $\alpha $ which
Lagrangian appeared due to the shift done in the auxiliary fields. Also, the
functional integral differential has been written as
\[
\mathcal{D}[\Phi ]=\mathcal{D}[A,\overline{\Psi },\Psi ,\overline{c},c],
\]
and the matrix $S^{-1}$ has the block structure
\begin{equation}
S^{-1}=\left\{
\begin{array}{lll}
\mathbf{A}/2 & \mathbf{C} & 0 \\
\mathbf{D} & \mathbf{B} & 0 \\
0 & 0 & \mathbf{G}
\end{array}
\right\} ,  \label{S}
\end{equation}
where the matrices \textbf{A}, \textbf{B}, \textbf{C}, \textbf{D }and\textbf{%
\ G }are defined by the expressions \textbf{\ }
\begin{eqnarray*}
\mathbf{A}^{(\mu ,a),(\nu ,b)}(\alpha ) &\equiv &-(\mathbf{D}^{ac}\mathbf{D}%
^{cb}\delta _{\mu \nu }-\frac{\mathbf{D}_\mu ^{ac}\mathbf{D}_\nu ^{cb}%
\mathbf{+D}_\nu ^{ac}\mathbf{D}_\mu ^{cb}}2+\frac{\delta ^{ab}}\alpha k_\mu
k_\nu ), \\
\mathbf{B}^{(u,r),(v,s)}(\alpha ,\overline{\chi },\chi ) &\equiv &\text{ }%
\gamma _\mu \text{(}k_\mu \delta ^{ij}+g\beta _\mu ^a\,T_a^{ij}), \\
\mathbf{D}^{(u,r),(\nu ,b)}(\overline{\chi },\chi ) &\equiv &-g(\frac{C_q^b}{%
(2\pi )^D})^{\frac 12}\gamma _v^{u\text{ }q}T_b^{ij}\chi ^{q,t}, \\
\mathbf{C}^{(\mu ,a),(v,s)}(\overline{\chi },\chi ) &\equiv &-g(\frac{C_q^b}{%
(2\pi )^D})^{\frac 12}\overline{\chi }^{q,t}\gamma _\mu ^{\text{ }q\text{ }%
v}T_b^{ts}, \\
\mathbf{G}^{ab}(\alpha ) &=&k_\mu \mathbf{D}_\mu ^{ab}, \\
\mathbf{D}_\mu ^{ab} &=&k_\mu \delta ^{ab}+gf^{abc}\text{ }\beta _\mu ^c, \\
\beta _\mu ^a &=&(\frac{2C_g^b}{(2\pi )^D})^{\frac 12}\alpha _\mu ^a.
\end{eqnarray*}

In expression (\ref{Zo}) the function of the auxiliary fields $\mathcal{M}$
is given by
\begin{eqnarray*}
\mathcal{M(}\alpha ,\overline{\chi },\chi ) &=&\frac 1{\mathcal{N}}\exp
[-\frac 14V^DF_{\mu \nu }^a(\alpha )F_{\mu \nu }^a(\alpha )]\times  \\
&&\exp [\int \frac{dk}{(2\pi )^D}\Phi ^{*}(-k)S^{-1}(k)\Phi (k) \\
&=&\frac 1{\mathcal{N}}\exp [-\frac 14V^DF_{\mu \nu }^a(\alpha )F_{\mu \nu
}^a(\alpha )]\times  \\
&&\int \mathcal{D}[A,\overline{\Psi },\Psi ,\overline{c},c]\exp [\int \frac{%
dk}{(2\pi )^D}(A(-k)\frac{\mathbf{A}(k)}2\text{ }A(k)+\overline{\Psi }(-k)%
\text{ }\mathbf{B}(k)\text{ }\Psi (k)+ \\
&&A(-k)\text{ }\mathbf{C}(k)\text{ }\Psi (k)\text{ +}\overline{\Psi }(-k)%
\text{ }\mathbf{D}(k)\text{ }A(k)+\overline{c}(-k)\mathbf{G}(k)c(k))] \\
&=&\frac 1{\mathcal{N}}\exp [-\frac 14V^DF_{\mu \nu }^a(\alpha )F_{\mu \nu
}^a(\alpha )]\times  \\
&&\text{ }Det^{-\frac 12}[\mathbf{A]}\text{ }Det[\mathbf{B-DA}^{-1}\mathbf{C]%
}Det[\mathbf{G]} \\
&=&\frac 1{\mathcal{N}}\text{ }\exp [-\frac 14V^DF_{\mu \nu }^a(\alpha
)F_{\mu \nu }^a(\alpha )]\times  \\
&&\exp [-\frac 12Tr[Log[\mathbf{A}]]+Tr[Log[\mathbf{B-DA}^{-1}\mathbf{C}%
]+Tr[Log[\mathbf{G}]]],
\end{eqnarray*}
in which the matrix indices of the fields with the block matrices had not
been written explicitly to avoid a more cumbersome expression. However,
their restitution seems to be clearly feasible. Henceforth, the following
expression can be written for the generating functional $Z$
\begin{eqnarray}
Z[j,\eta ,\overline{\eta },\xi \overline{,\xi }] &=&\frac{I[j,\eta ,%
\overline{\eta },\xi \overline{,\xi }]}{I[0,0,0,0]}  \label{finalZ} \\
I[j,\eta ,\overline{\eta },\xi ,\overline{\xi }] &=&\frac 1{\mathcal{N}}\int
\int d\alpha d\overline{\chi }d\chi \exp [-\overline{\chi }_u^i\chi _u^i-%
\frac{\alpha _\mu ^a\alpha _\mu ^a}2]\times \mathcal{M(}\alpha ,\overline{%
\chi },\chi )  \nonumber \\
&&\exp [\widetilde{V}^{int}[\frac \delta {\delta j},\frac \delta {\delta
\overline{\eta }},\frac \delta {-\delta \eta },\frac \delta {\delta
\overline{\xi }}\frac \delta {-\delta \xi },\alpha ,\overline{\chi },\chi ]]%
{\small \times }  \nonumber \\
&&\exp [\int \frac{dk}{(2\pi )^D}J^{*}(-k)S(k)J(k)],
\end{eqnarray}
in which the composite sources $J$ and propagator $S$ where defined in (\ref
{sources}) and (\ref{S}). The vertex terms $\widetilde{V}^{int}$ in (\ref
{finalZ}) are equal to the ones in $\widehat{V}^{int}$ plus one additional
second order in $g$ one which is linear in the gluonic auxiliary field. Its
expression is given at the end of Appendix A.

\subsubsection{Expansion parameters}

At this point is useful to recall that the modified perturbative expansion
under consideration has a set of three parameters on which the physical
quantities depend: ($g$,$C_q$,$C_g$). All of them will be assumed here to
have a dimension defined in powers of the renormalization scale $\mu.$
\,However, the alterative representation (\ref{finalZ}) suggests a
modification of the relevant parameters for the expansion in seeking for the
realization of the dimensional transmutation effect \cite{coleman}. This
idea comes from the form of the propagator $S$ in (\ref{finalZ}). It can be
noted that this new free propagator can be made gauge coupling independent
simply by defining the new set of independent parameters:
\begin{eqnarray}
g &=&g,  \nonumber \\
m_g^2 &=&g^2C_g,  \label{mass} \\
m_q^2 &=&\left( g^2C_q\right) ^{\frac 23}.  \nonumber
\end{eqnarray}

With this definition, the propagator of the composite field becomes coupling
independent and all the vertices non associated to counterterms are mass
independent. It can be remarked that these gluon and quark mass parameters
where the ones defining the prediction for the constituent masses in \cite
{epjc,jhep}.

It seem useful to resume here the dimensions of the various fields and
constants
\begin{eqnarray*}
D[A] &=&\frac{D-2}2,\ \;D[\Psi ]=\frac{D-1}2=D[\overline{\Psi }],\;\;D[\chi
]=\frac{D-2}2=D[\overline{\chi }], \\
D[g] &=&2-\frac D2,\;\;D[C_g]=D-2,\;\;\;D[C_q]=D-1.
\end{eqnarray*}
Then, the quark mass parameters $m_q$ having the expression $%
m_q^2=(g^2C_q)^{\frac 23}$ has a dimension equal to two not changing under
the regularization since
\[
D[m_q^2]=\frac 23(4-D+D-1)=2.
\]
The same is valid for the gluon mass parameter which dimension is
\[
D[m_g^2]=(4-D+D-s)=2.
\]

\subsubsection{Connected green Functions generator and Effective Action}

The connected Green functions generating functional $W$ and the Effective
Action $\Gamma $ are defined now by the usual Legendre transformation as
\begin{eqnarray*}
W[j,\eta ,\overline{\eta },\xi \overline{,\xi }] &=&\log [Z[j,\eta ,%
\overline{\eta },\xi ,\overline{\xi }], \\
\Gamma [A,\overline{\Psi },\Psi ,\overline{\chi },\chi ] &=&W[j,\eta ,%
\overline{\eta },\xi ,\overline{\xi }]-\int dx(\,j\,A+\overline{\eta }\Psi +%
\overline{\Psi }\eta +\overline{\xi }\chi +\overline{\chi }\xi ),
\end{eqnarray*}
in which the mean fields are determined from $Z$ through
\begin{eqnarray*}
A_\mu ^a(x) &=&\frac \delta {\delta j_\mu ^a(x)}\log (Z[j,\eta ,\overline{%
\eta },\xi ,\overline{\xi }]), \\
\Psi (x) &=&\frac \delta {\delta \overline{\eta }(x)}\log (Z[j,\eta ,%
\overline{\eta },\xi ,\overline{\xi }]),\overline{\Psi }(x)=\frac \delta
{\delta (-\eta (x))}\log (Z[j,\eta ,\overline{\eta },\xi ,\overline{\xi }]),
\\
c(x) &=&\frac \delta {\delta \overline{\xi }(x)}\log (Z[j,\eta ,\overline{%
\eta },\xi ,\overline{\xi }]),\text{ }\overline{c}(x)=\frac \delta {\delta
(-\xi (x))}\log (Z[j,\eta ,\overline{\eta },\xi ,\overline{\xi }]).
\end{eqnarray*}
It is important to notice that by the definition of the original $Z$, $W$ is
exactly the sum of all connected diagram in which the lines are the addition
of the usual Feynman propagators plus the ''condensate propagators''.
However, after the introduction of the auxiliary fields, the alternative
form of $Z$ became a mean value of generating functionals $Z(\alpha ,%
\overline{\chi },\chi )=\exp \,(\,W(\alpha ,\overline{\chi },\chi ))$
depending on the auxiliary fields. It can be suspected that the mean value
of $Z(\alpha ,\overline{\chi },\chi )$ coincides with the exponentiation of
the average of $W(\alpha ,\overline{\chi },\chi ),$ and therefore this
quantity should be equal with $W.$ However, we have not the proof of this
property yet. It will be considered in future extension of the work since
some hints point in the direction of its validity, at least approximately in
the infinite volume limit.

\section{Quark effective potential in order $g^0$}

Let us consider the one loop Effective Action when only the quark condensate
is retained. Then, $\Gamma $ at zero values of the mean fields and their
sources can be written in the form
\begin{eqnarray*}
\Gamma (m_q) &=&\log [\frac{Z[0,0,0,0,0|C_q,0]}{Z[0,0,0,0,0|0,0]}], \\
&=&\log {\LARGE [}\int \int d\overline{\chi }d\chi \exp (-\overline{\chi }%
\chi )\exp {\Large \{}V^{(D)}\int \frac{dk}{(2\pi )^D}Tr_{spin,color}{\Large %
(}\log [\mathbf{(}k_\mu \gamma _\mu \mathbf{)}^{(u,r),(v,s)}\mathbf{-} \\
&&\frac{m_q^3}{(2\pi )^D}\gamma _\mu ^{u\text{ }q}T_a^{rt}\chi ^{q,t}\frac
1{(-k^2)}\overline{\chi }^{q^{\prime },t^{\prime }}\gamma _\mu ^{\text{ }%
q^{\prime }\text{ }v}T_a^{t^{\prime }s}]-\log [\mathbf{(}k_\mu \gamma _\mu
\mathbf{)}^{(u,r),(v,s)}]{\Large )\}}{\LARGE ].}
\end{eqnarray*}

The mean value over the auxiliary field appearing above can be expressed as
follows
\begin{eqnarray*}
\mathcal{G} &=&\int \int d\overline{\chi }d\chi \exp (-\overline{\chi }\chi
)\exp {\LARGE \{}V^D\int \frac{dq}{(2\pi )^D}Tr_{spin,color}{\LARGE (}\log [%
\mathbf{(}q_\mu \gamma _\mu \mathbf{)}^{(u,r),(v,s)}\mathbf{-} \\
&&\gamma _\mu ^{u\text{ }q}T_a^{rt}\chi ^{q,t}\frac 1{(-q^2)}\overline{\chi }%
^{q^{\prime },t^{\prime }}\gamma _\mu ^{\text{ }q^{\prime }\text{ }%
v}T_a^{t^{\prime }s}]-\log [\mathbf{(}q_\mu \gamma _\mu \mathbf{)}%
^{(u,r),(v,s)}]{\LARGE )}{\Large \}} \\
&=&\int \int d\overline{\chi }d\chi \exp (-\overline{\chi }\chi )\exp
{\LARGE \{}V^D(\frac{m_q^3}{(2\pi )^D})^{\frac D3}\mathcal{F}(D)\}.
\end{eqnarray*}

The factor $\mathcal{F}(D)$ is only dependent on the dimension, and for this
pure quark case is convergent due to the high dimension of the parameter $%
m_q^3$. Its expression in terms of an integral over a dimensionless
variables takes the form
\begin{eqnarray}
\mathcal{F}(D) &=&\int \int d\overline{\chi }d\chi \exp (-\overline{\chi }%
\chi )\exp {\LARGE \{}\int \frac{dq}{(2\pi )^D}Tr_{D,C_f}{\LARGE (}\log [%
\mathbf{(}q_\mu \gamma _\mu \mathbf{)}^{(u,r),(v,s)}\mathbf{-}  \nonumber \\
&&\gamma _\mu ^{u\text{ }q}T_a^{rt}\chi ^{q,t}\frac 1{(-q^2)}\overline{\chi }%
^{q^{\prime },t^{\prime }}\gamma _\mu ^{\text{ }q^{\prime }\text{ }%
v}T_a^{t^{\prime }s}]-\log [\mathbf{(}q_\mu \gamma _\mu \mathbf{)}%
^{(u,r),(v,s)}]{\LARGE )}{\Large \}}.  \label{F}
\end{eqnarray}

It seems possible to evaluate $\mathcal{G}$ if the expansion over the
auxiliary fermion fields is properly investigated. However, this requires a
separate a detailed study to be considered elsewhere. Here, we will evaluate
it, in a kind of mean field approximation, in which the product of the
fields $\chi ^{q,t}\overline{\chi }^{q^{\prime },t^{\prime }}$ in (\ref{F})
will be replaced by its ''mean'' value over the integration of the auxiliary
fields. That is
\[
\chi ^{q,t}\overline{\chi }^{q^{\prime },t^{\prime }}->\int \int d\alpha d%
\overline{\chi }d\chi \exp (-\overline{\chi }\chi )\chi ^{q,t}\overline{\chi
}^{q^{\prime },t^{\prime }}=\delta ^{qq^{\prime }}\delta ^{tt^{\prime }}.
\]

After the use of the relations :
\begin{eqnarray*}
T_a^{ik}T_a^{kj} &=&C_F\text{ }\delta ^{ij},\;\;\;C_F=\frac{N^2-1}{2N}, \\
\gamma _\mu ^{u\text{ }q}\gamma _\mu ^{\text{ }q\text{ }v} &=&-D\text{ }%
\delta ^{uv},
\end{eqnarray*}
the $\mathcal{F}(D)$ factor gets the simple form
\begin{eqnarray*}
\mathcal{F}(D) &=&\exp {\LARGE \{}4N\int \frac{dq}{(2\pi )^D}\log [1+\frac{%
4D^2(N^2-1)^2}{4N^2}\frac 1{q^2}]{\LARGE \}} \\
&=&\exp {\LARGE \{}4N\frac{\pi ^{\frac D2}D}{\Gamma (\frac D2+1)}\int \frac{%
q^{D-1}dq}{(2\pi )^D}\log [1+\frac{4D^2(N^2-1)^2}{4N^2}\frac 1{(q^2)^3}]%
{\LARGE \}}.
\end{eqnarray*}

Therefore, the potential have the expression
\begin{eqnarray*}
V(m_q) &=&-\Gamma (m_q)=-V^D(\frac{g^2C_q}{(2\pi )^D})^{\frac D3}\mathcal{F}%
(D) \\
&=&-\frac{V^D}{(2\pi )^{\frac{D^2}3}}\mu ^4(\frac{m_q}\mu )^D4N\frac{\pi
^{\frac D2}D}{\Gamma (\frac D2+1)}\int \frac{q^{D-1}dq}{(2\pi )^D}\log [1+%
\frac{4D^2(N^2-1)}{4N^2}\frac 1{(q^2)^3}],
\end{eqnarray*}
which in the limit $D->4$ leads to the energy density $v(m_q)$
\[
v(m_q)=\frac{V(m_q)}{V^4}=-\frac{8\pi ^2N\text{ }m_q^4}{3(2\pi )^{\frac{16}3}%
}\int \frac{q^3dq}{(2\pi )^4}\log [1+\frac{64(N^2-1)^2}{4N^2}\frac
1{(q^2)^3}].
\]

\begin{figure}[tbp]
\begin{center}
\epsfig{figure=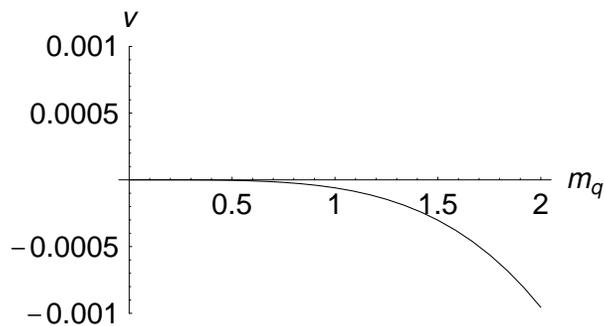,width=8cm} 
\end{center}
\caption{The energy density  estimated in the zeroth order in the
coupling approximation, plotted as a function of the quark mass parameter $%
m_q $. Note that in this low order approximation, the result indicates a
dynamical symmetry breaking under the generation of a quark condensate. The
same outcome was obtained in Ref. \onlinecite{epjc1} in which the same
result came from a less systematic analysis. The expansion parameters of any
quantity are assumed to be the coupling constant $g$ and the quark mass
parameter $m_q$}
\label{fig5}
\end{figure}

The dependence of $v(m_q)$ on the mass parameter $m_q$ is plotted in Fig. 1.
As in Ref. \cite{epjc2}, the result indicates a dynamical generation of the
quark condensate in this simple approximation. \quad The result is unbounded
from below. The possibility that higher order corrections could stabilize a
minimum will be investigated in future extensions of this work.

\section{Summay}

The renormalization of the modified perturbation expansion for mass less QCD
proposed in previous works is started to be investigated. A generating
functional $Z$ of the finite Green functions is constructed in terms of the
renormalized coupling, parameters and fields. Mass counterterms are also
introduced assumed their most probable need, since the theory is expected to
generate masses for the originally mass less fields. However, the bare
masses are assumed to vanish in consistency with the connection of the
interaction on a mass less theory which is reflected by the starting
unrenormalized generating functional. Expressions for the vertex terms are
given. The $Z$ functional is transformed to an alternative representation as
a mean value over a class of generating functionals associated to background
field theories in presence of constant and homogeneous auxiliary gluon and
quark fields. In this variant of the formulation, the gluon and quark field
are coupled in a global propagator mixing the boson and fermion fields. The
analysis suggests the convenience of introducing as the independent
expansion parameters for the physical quantities, the coupling constant $g$
as before, and two mass parameters $m_q$ and $m_g$. These parameters are
simply related with the quark and gluon condensate constants $C_q$ and $C_g$
respectively and retains their dimension equal to two under dimensional
regularization. Their relation to the constants reflecting the quark and
gluon condensates are $m_q^2=(g^2C_q)^{\frac 23\text{ }}$ and $m_g^2=g^2C_q.$
They were relevant in the prediction of the constituent quark masses done in
\cite{epjc,epjc2}.

An evaluation of the lowest order contribution to the Effective Action is
presented for the case in which only a quark condensate is retained. The
renormalization at this modified tree level was not required. This lower
order result indicates a dynamical generation of a quark condensate. This
prediction was also obtained in Ref. \onlinecite{epjc2}. However, here it is
appearing from a more systematic framework. For the extension of the work,
it is planned to investigate the possibilities that higher order
contributions could produce a minimum of the potential. Such a result might
open a way for the application of the modified expansion in justifying a
kind of Top condensate model as an effective description of mass less QCD.

\begin{acknowledgments}
The invitation and kind hospitality of the High Energy Section of
the Abdus Salam International Center for Theoretical Physics
(ASICTP) and its Head S. Randjbar-Daemi, allowing for a very
helpful visit to the Center, is deeply acknowledged. I express
also my gratitude by the support to the work received from the
Office of External Activities of ICTP (OEA),  through the Network
on \textit{Quantum Mechanics, Particles and Fields}(Net-35).  The
useful remarks during the stay at the AS ICTP received from G.
Thompson and K. Narain are also very  much appreciated.
\end{acknowledgments}

\appendix

\section{}
 The explicit form of all the counterterms that appear in
(\ref{vint}) is the following

\begin{eqnarray*}
V_g^{(1)}[\frac \delta {\delta j}] &=&\frac 1{3!}\int \int dk_1dk_2dk_3(2\pi
)^D\delta ^D(k_1+k_2+k_3)\times \\
&&V_{\mu _1\mu _2\mu _3}^{a_1a_2a_3}(k_1,k_2,k_3)\frac \delta {\delta j_{\mu
_1}^{a_1}(k_1)}\frac \delta {\delta j_{\mu _2}^{a_2}(k_2)}\frac \delta
{\delta j_{\mu _3}^{a_3}(k_3)},
\end{eqnarray*}
\begin{eqnarray*}
V_g^{(2)}[\frac \delta {\delta j}] &=&\frac 1{4!}\int \int
dk_1dk_2dk_3dk_4(2\pi )^D\delta ^D(k_1+k_2+k_3+k_4)\times \\
&&V_{\mu _1\mu _2\mu _3\mu _4}^{a_1a_2a_3a_4}(k_1,k_2,k_3,k_4)\frac \delta
{\delta j_{\mu _1}^{a_1}(k_1)}\frac \delta {\delta j_{\mu
_2}^{a_2}(k_2)}\frac \delta {\delta j_{\mu _3}^{a_3}(k_3)}\frac \delta
{\delta j_{\mu _4}^{a_4}(k_4)},
\end{eqnarray*}
\begin{eqnarray*}
V_q^{(1)}[\frac \delta {\delta \overline{\eta }},\frac \delta {-\delta \eta
},\frac \delta {\delta j}] &=&\int \int dk_1dk_2dk_3(2\pi )^D\delta
^D(k_1+k_2+k_3) \\
&&\frac \delta {\delta \eta ^{i_1}(k_1)}\,g\,\gamma _{\mu
_3}T_{a_3}^{i_1i_2}\frac \delta {\delta \overline{\eta }^{i_2}(k_2)}\frac
\delta {\delta j_{\mu _3}^{a_3}(k_3)},
\end{eqnarray*}
\begin{eqnarray*}
V_{gh}^{(1)}[\frac \delta {\delta \overline{\xi }},\frac \delta {-\delta \xi
},\frac \delta {\delta j}] &=&\int \int dk_1dk_2dk_3(2\pi )^D\delta
^D(k_1+k_2+k_3) \\
&&\frac \delta {\delta \eta ^{a_2}(k_2)}\,(-ig)k_{2\mu
_3}\,f^{a_1a_2a_3}\frac \delta {\delta \overline{\xi }^{a_3}(k_3)}\frac
\delta {\delta j_{\mu _3}^{a_3}(k_3)},
\end{eqnarray*}
\begin{eqnarray*}
V_g^{(0)}[\frac \delta {\delta j}] &=&\frac 1{3!}\int \int dk_1dk_2(2\pi
)^D\delta ^D(k_1+k_2)\times \\
&&\frac \delta {\delta j_{\mu _2}^{a_2}(k_2)}\frac{(Z_3-1)}2\delta
^{a_2a_1}(k_1^2\delta _{\mu _2\mu _1}-k_{1\mu _2}k_{1\mu _1})\frac \delta
{\delta j_{\mu _1}^{a_1}(k_1)},
\end{eqnarray*}
\begin{eqnarray*}
V_q^{(0)}[\frac \delta {\delta \overline{\eta }},\frac \delta {-\delta \eta
}] &=&\int \int dk_1dk_2(2\pi )^D\delta ^D(k_1+k_2) \\
&&\frac \delta {\delta \eta (k_2)}[(Z_2-1)k_{1\mu }\gamma _\mu +\delta
m(g_0,C_q^0,C_g^0)]\frac \delta {\delta \overline{\eta }(k_1)},
\end{eqnarray*}
\begin{eqnarray*}
V_{gh}^{(0)}[\frac \delta {\delta \overline{\xi }},\frac \delta {-\delta \xi
}] &=&\int \int dk_1dk_2(2\pi )^D\delta ^D(k_1+k_2) \\
&&\frac \delta {\delta \xi ^{a_2}(k_2)}(Z_3-1)\delta ^{a_2a_1}k_1^2\frac
\delta {\delta \overline{\xi }^{a_1}(k_1)}.
\end{eqnarray*}

The only new vertex that appears in $\widetilde{V}^{int}$ in (\ref{finalZ})
after the quadratic form in the gluon and quark fields depending on the
auxiliary fields is extracted in order to arrive to this expression (\ref
{finalZ}), has the form: .
\begin{eqnarray*}
V_g^{(2)}[\frac \delta {\delta j}] &=&\frac 1{4!}\int \int
dk_1dk_2dk_3dk_4(2\pi )^D\delta ^D(k_1+k_2+k_3+k_4)\times \\
&&V_{\mu _1\mu _2\mu _3\mu _4}^{a_1a_2a_3a_4}(k_1,k_2,k_3,k_4)(\frac \delta
{\delta j_{\mu _1}^{a_1}(k_1)}+4(\frac{2C_g}{(2\pi )^D})^{\frac 12}\alpha
_{\mu _1}^a\delta ^D(k_1))\frac \delta {\delta j_{\mu _2}^{a_2}(k_2)}\frac
\delta {\delta j_{\mu _3}^{a_3}(k_3)}\frac \delta {\delta j_{\mu
_4}^{a_4}(k_4)}.
\end{eqnarray*}

\end{document}